\documentstyle[12pt]{article} \textwidth 6in \textheight 8.5in
\def\be{\nopagebreak[3]\begin{equation}}
\def\ee{\end{equation}}
\def\ba{\nopagebreak[3]\begin{eqnarray}}
\def\ea{\end{eqnarray}}

	\def\ni{\noindent}
	
\begin{document}
	\begin{center}
	\centerline{}
	\vspace{.5cm}
	{\LARGE\bf Probing Quantum Gravity Through Exactly
	Soluble Midi-Superspaces I\\}
	\vspace{0.7cm}
	{\large\em A. Ashtekar and M. Pierri\\}
	\vspace{0.5cm}
	{Center for Gravitational Physics and Geometry \\
	Department of Physics, The Pennsylvania State University  \\
	University Park, PA 16802\\ }
	\vspace{0.3cm}
	{\small CGPG-96/5-3\\ }
	\vspace{0.5cm}
	\end{center}

\begin{abstract}

It is well-known that the Einstein-Rosen solutions to the 3+1
dimensional vacuum Einstein's equations are in one to one
correspondence with solutions of 2+1 dimensional general relativity
coupled to axi-symmetric, zero rest mass scalar fields. We first
re-examine the quantization of this midi-superspace paying special
attention to the asymptotically flat boundary conditions and to
certain functional analytic subtleties associated with regularization.
We then use the resulting quantum theory to analyze several conceptual
and technical issues of quantum gravity.

\end{abstract}

\section{Introduction}
\label{s1}

Many of the central problems of quantum gravity can be traced back to
two main difficulties: i) the absence of a background space-time
metric; and, ii) the presence of an infinite number of degrees of
freedom.

Let us begin with the first set of issues.  The absence of a
background geometry implies that the theory has to be diffeomorphism
invariant and this feature makes it difficult to construct observables
and formulate precisely questions of direct physical interest. It also
gives rise to the celebrated ``problem of time'': if there is no
background metric, what are we to make of the notion of ``time
evolution''? Indeed, if the diffeomorphisms are to be regarded as
gauge, at first sight, dynamics also appears as a part of gauge. Can
one disentangle dynamics from gauge unambiguously?
These questions are of course not new. (For a detailed discussion, see,
e.g. \cite{as}.) To gain insight in to these issues, a number of
mini-superspace models have been discussed in the literature (see,
e.g., \cite{mini}). In Bianchi models, for example, one restricts
attention only to spatially homogeneous solutions of Einstein's
equations and, in the quantum theory, addresses the issue of time via
``de-parametrization''. Perhaps a more striking model is presented by
2+1-dimensional vacuum general relativity which, like the 3+1 theory,
is fully diffeomorphism invariant. Quantization of this model
\cite{sc:94, aa:91} has shed light on the notion of observables, role of
discrete symmetries, etc. These models have also given us considerable
insights into the technical problems that arise due to the underlying
diffeomorphism invariance.  For example, since we have no Poincar\'e
group to help us, the problem of finding the correct inner-product on
the space of quantum states requires a new strategy. The 2+1 model has
provided a method which, moreover, is free of ambiguities that arise,
e.g., in the de-parametrization procedure.

However, these models do not come to grips with the second main
difficulty mentioned above: the presence of an infinite number of
degrees of freedom. To face this difficulty, we need to consider
genuine field theories which do not require a background space-time
metric. An obvious strategy would be to again consider symmetry
reductions which, however, are mild enough to leave behind {\it local}
degrees of freedom. To locate convenient choices, let us briefly
return to the 2+1 dimensional vacuum general relativity. This theory
can be obtained by a symmetry reduction of 3+1-dimensional general
relativity with respect to a single space-like Killing field
which is hyper-surface orthogonal and whose norm is {\it
constant}. Therefore, as a next step, it is natural to drop the severe
condition on the norm. The symmetry reduced system now has an infinite
number of degrees of freedom. In fact it is now equivalent to
2+1-dimensional general relativity coupled to a zero rest mass scalar
field (which is given by the logarithm of the norm of the Killing
field) \cite{krammer, aamv94}. Unfortunately, this midi-superspace is
a bit too complicated in that the issue of global existence of such
solutions is still largely unexplored in the classical
theory. However, if we require that there be another hyper-surface
orthogonal Killing field in the 3+1 theory which commutes with the
first one, the situation simplifies dramatically. For, now one can in
effect ``decouple'' gravity and the scalar field. More precisely, the
equation satisfied by the scalar field on the curved 2+1 dimensional
space-time is equivalent to the wave equation on a fictitious {\it
flat} 2+1-dimensional space-time. Therefore, one can first solve the
second equation without any reference to the physical metric and then
use the solution to obtain the physical metric by simple
integration. Classically, one now has complete control on the issue of
global existence.

Such space-times were considered by Einstein and Rosen in the thirties
for the case when the first Killing field is a translation in the
``z-direction'' and the second is a ``x-y rotation''.  Thus, they
represent cylindrical gravitational waves (with only one polarization
because of the hyper-surface orthogonality requirement.) Their
quantization was considered in a remarkable paper by Kucha\v{r}
\cite{kk71} already in 1971. The problem was considered again from a
2+1-dimensional perspective by Allen \cite{ma87} in 1987 (without,
however, realizing that this is precisely a symmetry reduced version
of \cite{kk71}.) In the present paper, we shall return to this
midi-superspace. Our purpose is two-folds: i) to supplement the
analyses by Kucha\v{r} and Allen with a careful treatment of boundary
conditions in the classical theory and of certain functional analytic
issues in the quantum theory; and, ii) to use the resulting quantum
theory to analyze several conceptual and technical problems of quantum
gravity. Since the model itself is simple enough to be exactly
soluble, it provides a concrete arena to examine these vexing issues
and to see how they can be resolved in practice.

Specifically, following \cite{kk71, ma87}, we will use a canonical
approach. Since in this approach one {\it begins} with a 2+1
decomposition of space-time, apriori it is not clear if quantized {\it
space-time} geometries can emerge in the final theory. Indeed, one
often hears the criticism that, since it is tied to space-like
surfaces, the canonical approach may be inadequate to handle
``space-time issues'' such as ``fluctuations of the light
cone''. Here, we have a complete quantum theory. It is therefore
natural to ask: are there operators on the final Hilbert space
corresponding to space-time geometries? If so, is there adequate
structure to analyze how the light cones fluctuate?  More generally,
can we tie the canonically quantized theory to the quantum description
that emerges from covariant approaches? Can we compute $S$-matrices?
In the classical theory, there is a positive energy theorem
\cite{aamv94, mv95}. Does it continue to hold in the quantum theory?
Is the true ground state ``peaked around'' Minkowski space-time?  Or,
does the ground state contain wild quantum fluctuations with Planck
energy density as suggested by Wheeler \cite{jw:63}?  If so, the true
ground state would not have much resemblance to Minkowski space,
except perhaps on a suitable coarse-graining. Another question which
plays an important role in semi-classical considerations is: Are there
``coherent states'' which are peaked at classical solutions?

There is a non-perturbative approach to full quantum gravity which is
based on connections and triads (see, e.g., \cite{aa:91}). A basic
assumption in that approach is that the ``Wilson-loop operators''
--which correspond to traces of holonomies of a connection around
space-like loops-- should be well-defined. Apriori it is not clear if
this assumption is a reasonable one since in the definition of these
operators, one appears to smear a quantum field along a {\it one}
dimensional object (rather than three or four). It is natural to ask
for the status of this assumption in a completely solved model. Are
these Wilson loop operators well-defined on the explicitly known
quantum Hilbert space?  

Of course, just because such questions are answered in one way in a
specific solution to this model, does not imply that they are not
answered in another way in another solution and, more importantly, in
full 3+1-dimensional quantum gravity. Nonetheless, the ability to answer
them in detail in an explicitly solution can contribute substantially
to our overall intuition for quantum gravity. Our analysis is
primarily motivated by such considerations. We will find that most of
these questions can be answered in detail but that the analysis
involves several rather subtle points.

The plan of the paper is as follows. In section 2, we consider the
classical Hamiltonian formulation and isolate the true degrees of
freedom by a gauge fixing procedure. Because we are in an
asymptotically flat situation, by treating the boundary conditions
carefully, we can distinguish gauge from dynamics. In particular, the
true degrees of freedom are naturally subject to non-trivial dynamics
(without the need of any ``deparametrization''.) In section 3, we
calculate the classical Wilson loop functions and express them in
terms of the true degrees of freedom. Quantization is taken up in
section 4. As in \cite{kk71,ma87} the Hilbert space of states is a
Fock space for scalar fields in 2+1 dimensions. Subtleties arise,
however, because the geometrical observables --such as the space-time
metric and the Wilson loops-- are expressed as integrals of quadratic
functionals of these elementary excitations. Thus, in a rough
terminology, geometric excitations arise as non-local ``collective
modes'' of the primary mathematical entity, the quantum scalar field.
Finally, questions raised earlier in this section are analyzed within
this solution. Section 5 summarizes the main results and points out
directions for further work.
 
\section{Hamiltonian Formulation}
\label{s2}

\subsection{The midi-superspace}\label{s2.1}

Let us begin with a precise specification of our midi-superspace. For
definiteness, we will work in the 2+1-dimensional formulation. Thus,
we will consider asymptotically flat, axi-symmetric solutions of
2+1-dimensional general relativity coupled to zero rest mass
scalar-fields (where the rotational Killing field is hyper-surface
orthogonal). The underlying manifold $M$ will be topologically $R^3$
and the space-time metric will have signature --,+,+. For simplicity,
we will assume that all fields under consideration are $C^\infty$.

Denote by $\sigma^a$ the rotational Killing field. Hyper-surface
orthogonality of $\sigma^a$ implies that the space-time metric
$g_{ab}$ has the form:
\be
g_{ab} = h_{ab} + R^2\,{\nabla_a \sigma\, \nabla_b\sigma}
\label{0.1}
\ee
where $R$ is the norm of the Killing field and $\sigma$ is the
``angular coordinate''; $\nabla_a \sigma = R^{-2}g_{ab}\sigma^b$. The
field $h_{ab}$ so defined is a metric of signature --, + on the
2-manifolds orthogonal to $\sigma^a$.  Let us introduce a space-like
foliation of this 2-manifold by lines $t ={\rm const}$ and a dynamical
vector field $t^a = Nn^a+ N^r \hat{r}^a$, where $n^a$ is the unit,
time-like normal to the foliation and $\hat{r}^a$ the unit (outgoing)
vector field within each slice.  The pair $N, N^r$ constitutes the
lapse and the shift. If we now introduce a radial coordinate $r$ on
any one leaf such that $r=0$ at the axis (i.e., where $R =0$) and $r$
tends to infinity at spatial infinity, the 2-metric $h_{ab}$ can
be written as:
\be 
h_{ab} = (-N^2 + (N^r)^2) \nabla_a t\, \nabla_b t + 2N^r \nabla_{(a} t\,
\nabla_{b)}r + e^\gamma \nabla_a r\, \nabla_b r\, ,
\label{0.2}
\ee
where $N, N^r$ and $\gamma$ are functions of $r$ and $t$. It is
because of axi-symmetry, that the 3-metric $g_{ab}$ has only four
independent components and they are functions only of two variables.

Thus, our midi-superspace consists of five functions, $(N, N^r,
\gamma, R, \psi)$ on the space-time manifold $M$ where $\psi$ is the
zero rest mass scalar field (which is also Lie-dragged by the
rotational Killing field). The five fields are subject to the
following field equations:
\be
G_{ab} = T_{ab} \quad{\rm and}\quad 
g^{ab}\nabla_a\nabla_b \psi = 0\, ,
\ee
where $G_{ab}$ is the Einstein tensor of $g_{ab}$ which is determined
by the fields $(N, N^r, \gamma, R)$ via (\ref{0.2}) and $T_{ab}$ is
the stress-energy tensor of the scalar field $\psi$:
\be
T_{ab} = \nabla_a \psi\, \nabla_b \psi - {\textstyle{1\over 2}} (g^{cd}
\nabla_c \psi \nabla_d \psi) g_{ab}\, .
\ee
(Here, we have used a normalization that arises naturally in the
reduction from the 3+1 theory to the 2+1. {}From the 2+1 perspective, it
is natural to regard $\phi := \psi/\sqrt{8\pi G}$ as the physical
Klein-Gordon field, where $G$ is Newton's constant.)

Asymptotic flatness and regularity at the axis imply certain boundary
conditions on our dynamical fields. We first note that $g_{ab}$
reduces to a Minkowskian metric when $N=1, N^r= 0, \gamma= 0, R= r$
and $\psi =0$. The general asymptotic flatness conditions can be
written as:
\ba
N &=& 1 + N_1(r,t),\quad\quad N^r = N^r_o(t) + N^r_1 (r,t)
\nonumber\\
\gamma(r,t) &=& \gamma_\infty(t) + \gamma_1(r,t),
\quad\quad R(r,t) = r (1+ R_1(r,t)) 
\ea
where, on any $t= {\rm const}$ surface, $N_1, N^r_1, \gamma_1, R_1$
and the scalar fields $\psi$ are of asymptotic order $O(1/r)$. (We
will say that a function $f(r)$ is of asymptotic order $1/r$ if
$rf(r), r^2f'(r)$ and $r^3f''(r)$ admits limits as $r$ tends to
infinity, where a prime denotes a derivative with respect to $r$.)
While the conditions imposed on $N, N^r,R$ and $\psi$ are the obvious
ones, the condition on the field $\gamma$ seems surprising at
first. For, even at infinity, $\gamma$ is not required to approach its
Minkowskian value, $0$. The reason is that the asymptotic value of
$\gamma$ contains the information about mass: If $\gamma_\infty \not=
0$, the spatial metric has a deficit angle at infinity which measures
the ADM mass \cite{aamv94, mv95}. Thus, there is a striking contrast
with asymptotic flatness in 3+1 dimensions; the space-time metrics in
our midi-superspace do {\it not} approach a fixed Minkowskian metric
at infinity. Note finally that these boundary conditions are somewhat
simpler than those used in \cite{mv95} where general 2+1-dimensional
space-times were considered. Here, we can exploit the fact that we are
now working in a highly restrictive context of cylindrical waves.

Finally, regularity at the axis is ensured by requiring that $N^r,
\gamma$ and $R$ vanish there for all $t$. (Recall also that by
assumption, $N, N^r, \gamma, R^2$ and $\psi$ are $C^\infty$ everywhere
and, in particular, at $r=0$.)

\subsection{Phase Space}\label{s2.2}

Let us begin with the 3-dimensional action with appropriate boundary 
terms:
\be 
S(g, \psi) := {1\over 16\pi G} \int_{M'} d^3x\, \sqrt{g} [{\cal R}  
- g^{ab}\nabla_a \psi\nabla_b \psi] + {1\over 8\pi G} 
\oint_{\partial M'} d^2x [K\sqrt{h} - K_o\sqrt{h_o}]\, ,
\label{action}
\ee
where $M'$ is an open set in $M$; $\partial M'$, its boundary in $M$;
${\cal R}$, the scalar curvature of $g$; $K$ and $h$, the trace of the
extrinsic curvature of, and the determinant of the intrinsic metric on
$\partial M'$ induced by $g_{ab}$; and, $K_o$ and $h_o$ are the
corresponding fields induced by the Minkowski metric $\stackrel{\circ}
{g}_{ab}$ (obtained by setting $N=1, N^r =0, \gamma= 0, R=r$ and
$\psi= 0$).

To pass to the Hamiltonian formulation, one performs a
2+1-decomposition. Let us substitute in (\ref{action}) the form of the
metric given by Eqs. (\ref{0.1}) and (\ref{0.2}). Then,
the action reduces to the standard form:
\be
S = {1\over 8G}\int dt\, \left( dr (p_\gamma \dot{\gamma} + p_R \dot{R}
+ p_\psi \dot{\psi})\,\, -\, H [N, N^r] \right)
\ee
The Hamiltonian $H$ is given by:
\be\label{ham}
H [N, N^r] = {1\over 8G}\int dr (NC + N^rC_r)\,\, + {1\over 4G} 
(1 - e^{- \gamma_\infty/ 2})
\ee
where $C$ and $C_r$ are functions of the canonical variables,
\ba\label{constraints}
C & = &  e^{-\gamma/2} (2R'' - {\gamma}'R' - p_{\gamma}p_R ) + 
\frac{1}{2} R \, e^{-\gamma/2} (\frac{{p_\psi}^2}{R^2} + {{\psi}'}^2),   
      \nonumber\\
C_r & = &  e^{-\gamma} ( -2 {p'}_{\gamma} + \gamma' p_{\gamma} + R'p_R ) 
+ e^{-\gamma} p_{\psi} \psi' \, ,
\label{7}
\ea
and $\gamma_\infty$ is the value of $\gamma$ at $r=\infty$.  (Here
primes denote derivatives with respect to $r$.)

As expected, the lapse and shift functions $N, N^r$ appear as Lagrange
multipliers; they are not dynamical variables.  Thus, the phase-space
$\Gamma$ consists of three canonically-conjugate pairs, $(\gamma,
p_\gamma; R, p_R; \psi, p_\psi)$, on a 2-manifold $\Sigma$ which is
topologically $R^2$. The boundary conditions on the configuration
variables $(\gamma, R, \psi)$ have already been discussed. The
conditions on the momenta can be deduced from their definitions in
terms of these fields and their time derivatives.  At infinity,
$p_\gamma$ and $p_R$ fall-off as $O(1/r^2)$ while $p_\psi$ falls off
as $O(1/r)$. (Note that these conditions imply that action $\int dr
p_\gamma \delta\gamma, \int dr p_R \delta R $ and $\int dr p_\psi
\delta\psi$ of the momenta on the tangent vectors $\delta \gamma,
\delta R, \delta\psi$ to our configuration space are all finite, so
that we have a well-defined (weakly non-degenerate) symplectic
structure.) There are two first class constraints, $C = 0$ and $C_r =
0$, obtained by varying the action with respect to the Lagrange
multipliers $N$ and $N^r$. The Hamiltonian is given by $H$. (It is
because of the underlying axi-symmetry that we have only one
diffeomorphism constraint, $C_r$.)

Let us begin by analyzing the canonical transformations generated by
constraints. For this, we have to first smear the constraints and
obtain well-defined functions on the phase space, say, $C[N_g]:=\int
dr N_gC$ and $C[N_g^r]= \int dr N_g^r C_r$. Using the boundary
conditions on the phase space variables, it is straightforward to
verify that these functions are well-defined {\it and} differentiable
on the phase space if $N_g$ vanishes on the axis and is of asymptotic
order $O(1/r)$ and $N_g^r$ admits a limit at infinity. ({}From now on,
the subscript $g$ on smearing fields will indicate that they satisfy
these boundary conditions.) Since the constraints are of first class,
and since we are in the asymptotically flat context, the canonical
transformations generated by these constraints can be regarded as
``gauge'' in an appropriate sense. As one might expect, $C[N_g]$
generates ``bubble time evolutions'' via lapses which go to zero at
infinity while $C[N_g^r]$ generates spatial diffeomorphisms which are
bounded at infinity. The situation with the Hamiltonian constraint is
the same as the one we normally encounter in the 3+1-dimensional
theory. For the diffeomorphism constraint, on the other hand, the
situation is quite different since the diffeomorphisms generated by
$N_g^r \hat{r}^a$ are not necessarily asymptotically identity. This
is, however, the standard situation in 2+1 dimensions (see e.g.,
\cite{aamv94, mv95}): In 2+1 dimensions, there are no asymptotic
Killing fields corresponding to spatial translations and the ADM
2-momentum vanishes.

To obtain genuine time translations, we have to allow lapses which
tend to $1$ at infinity and on the axis. When this is done, the
constraint function $C[N]$ continues to exist everywhere on the phase
space. However, due to the presence of the first two terms involving
derivatives of $\gamma$ and $R$ in the expression of $C$, the function
$C[N]$ fails to be differentiable. To make it differentiable, one has
to add a surface term.  As one might expect, this is precisely the
surface term in the expression (\ref{ham}) of the Hamiltonian. Thus,
the function which generates the canonical transformation
corresponding to (asymptotically unit) time translation is precisely
the Hamiltonian $H [N]$ (obtained by setting $N^r =0$ in
Eq. (\ref{ham})). On physical states --i.e., when the constraints are
satisfied-- the numerical value of the Hamiltonian is given by the
surface term in (\ref{ham}):
\be \label{energy}
E= {1\over 4G} (1-e^{-{1\over 2}\gamma_\infty}),  
\ee
As usual, in the space-time picture, the evolution generated by the 
Hamiltonian on the phase space corresponds to motions along the 
vector field $t^a$.

Let us summarize the discussion of this sub-section. Because we are in
the asymptotically flat context, there is a clean separation between
gauge and dynamics. As usual, when it comes to physical
interpretation, the ``gauge transformations'' of general relativity
have a somewhat different status from that in Yang-Mills theory. It is
not that the diffeomorphisms generated by $C[N_g]$ and $C[N^r_g]$ are
``unphysical''. Rather, they are ``redundant'' when it comes to
extracting the physical content of the theory. As we will see below,
we can gauge fix these constraints and extract the true degrees of
freedom. The Hamiltonian generates ``time evolution'' among these
gauge fixed points. Knowing this evolution, we can reconstruct the
entire solution; motions generated by constraints are not needed and
are in this sense ``redundant''.

\subsection{Gauge Fixing}\label{s2.3}

Since the canonical transformations generated by $C[N_g]$ and
$C[N^r_g]$ are to be regarded as gauge, as in Yang-Mills theory, to
gauge fix the system we need to extract one point from each orbit of
the corresponding Hamiltonian vector fields. This is achieved by
imposing gauge fixing conditions which, together with the constraints,
constitute a second class system. As in \cite {kk71}, we will choose
these conditions to make the space-time geometry transparent.  Let us
demand:
\be \label{gfc}
R(r)= r \quad{\rm and} \quad  p_\gamma (r) =  0 \, . 
\label{11}
\ee
\noindent
The first condition is motivated by the fact that, in any solution to
the field equations (satisfying our boundary conditions), the gradient
$\nabla_a R$ of the norm of the Killing field
$\partial /\partial \sigma$ is space-like everywhere on $M$
\cite{bbetal94}.  Since furthermore $R \sim r$ at the axis and at
infinity, it is natural to use $R$ itself as the radial
coordinate. After this condition is imposed, $R$ will no longer be a
dynamical variable. The second gauge fixing condition will remove
$\gamma$ from our list of dynamical variables. Thus, if these
conditions are admissible, the true degrees of freedom will all reside
in the field $\psi$, in accordance with our general expectation that
in 2+1 dimensions, all the local degrees of freedom are carried by
matter fields.
 
To see if our gauge fixing conditions are admissible, let us compute
their Poisson brackets with the constraints. We have:
\ba
\{R(r)-r , C_r[N^r_g]\}  &=&  N^r_g e^{-\gamma}\, R' \nonumber\\
\{p_\gamma , C[N_g]\}  &=&    \left[\frac{N_g}{2}\right. 
\left(-p_\gamma p_R + \frac{p^2_\psi}{2R} + \right.
    \left. \left. \frac{R}{2} {\psi'}^2 \right)  - {N_g}' R' \right] 
    e^{-{\gamma\over 2}}\, ,
\label{12}
\ea
where, as before, $N^r_g$ and $N_g$ are pure gauge lapses and shifts.
If $N_g^r\not= 0$ and $N_g \not= 0$, the right sides of (\ref{12}) do
not vanish at any point on the intersection of the surfaces defined by
constraints and gauge fixing conditions (\ref{gfc}).  Hence, as
needed, the gauge fixed surface intersects the gauge orbits
transversely.

The question now is whether we can choose lapse and shift such that
the dynamical evolution generated by the Hamiltonian $H[N, N^r]$
preserves the gauge conditions. More precisely, since the Hamiltonian
$H[N, N^r]$ weakly commutes with the constraints $C[N_g], C[N^r_g]$,
we know that the dynamical evolution it generates maps entire gauge
orbits to entire gauge orbits. The question is if we can select $N,
N^r$ such that the image under evolution of any gauge fixed point on
the constraint surface is another gauge fixed point. General
considerations from symplectic geometry imply that if such a pair
exists, it is unique. We will now establish the existence. Let us
begin with the Poisson brackets between the gauge conditions and the
Hamiltonian:
\ba
\{R(r)-r, H[N,N^r]\} & \approx &  N^r e^{-\gamma}
\nonumber \\
\{p_\gamma (r), H[N,N^r]\} & \approx  &   \left[\frac{N}{4r}\right. 
    \left(p^2_\psi + \right.
    \left. \left. {r^2} {\psi'}^2 \right)  - {N}' \right] 
    e^{-{\gamma\over 2}}\, ,
\label{13}
\ea
where $\approx$ stands for equality modulo constraints and gauge
conditions. We seek $N$ and $N^r$ which satisfy our boundary
conditions (namely, $N = 1 + O(1/r)$ and $N^r = N^r_o+O(1/r)$ at
infinity) and for which the right hand sides of (\ref{13}) vanish
(modulo constraints and gauge conditions). The only solutions are:
\be 
N(r) = \exp\, -{\textstyle{1\over 4}} \int_r^\infty\, dr_1 r_1
\left({(p_\psi)^2\over r_1^2} + (\psi')^2 \right) \quad
\hbox{and} \quad
N^r(r) =0.
\label{14}
\ee

Finally, let us extract the true degrees of freedom of the theory.  In
order to accomplish this, we need to eliminate redundant variables by
solving the set of second class constraints (\ref{7}) and use gauge
conditions (\ref{11}). By setting $R =r$ and $p_\gamma = 0$ in
(\ref{7}), we can trivially solve for $\gamma$ and $p_R$ in terms of
$\psi$ and $p_\psi$ (using the Hamiltonian and the diffeomorphism
constraints respectively). The result is:
\ba 
\gamma(R) & = & \frac{1}{2} \int_0^R dR_1
R_1\left(\frac{{p_\psi}^2} {R_1^2} + \right. \left. {\psi'}^2
\right), \label{15}\\ 
p_R & = & -p_\psi \psi' \label{16} 
\ea
Substituting (\ref{15}) in (\ref{14}), we can also express the lapse
$N$ in terms of $\gamma$. Thus, as expected, the true degrees of
freedom reside just in the matter variables. Indeed, the space-time
metric is now completely determined by $\psi$ and $p_\psi$:
\be 
\label{17} g_{ab} = e^{\gamma(R,t)}\left(-
e^{-\gamma_\infty\,} \nabla_a t\, \nabla_b t\, + \nabla_a R
\nabla_b R \right) + R^2 \nabla_a\sigma\, \nabla_b\sigma\, , 
\ee
where, from now on, $\gamma$ will only serve as an abbreviation for
the right side of (\ref{15}).

\subsection{Reduced Phase Space}\label{s2.4}

It is obvious from the above discussion that the reduced phase space
$\bar\Gamma$ can be coordinatized by the pair $(\psi(R), p_\psi
(R))$. The (non-degenerate) symplectic structure on the reduced phase
space $\bar\Gamma$ is the pull-back of the symplectic structure on
$\Gamma$. Thus,
\be
\{\psi (R_1) , p_\psi (R_2) \} = \delta(R_1 , R_2) 
\label{14c}
\ee
on $\bar\Gamma$. Next, let us write the reduced action by substituting
(\ref{11}), (\ref{15}) and (\ref{16}) in (\ref{action}),
\be
S[\psi,p_\psi] = \frac{1}{8G} \int dt \left[\int dR (p_\psi {\dot\psi} )\,  
\right. \left. - 2 (1 - e^{-{{1\over 2}\gamma_\infty }}) \right],
\label{19}
\ee
where, as before, $\gamma_\infty = \gamma (r\!=\!\infty)$.  By varying
the action (\ref{19}) with respect to $\psi$ and $p_\psi$ we then
obtain equations of motion:
\be \label{eom}
\dot\psi = e^{-{1\over 2}\gamma_\infty}\, {p_\psi\over R} 
\quad{\rm and}\quad \dot{p}_\psi = 
e^{-{1\over 2}\gamma_\infty}\, (R\psi')' \, .
\ee
Due to the presence of $\exp (-{\gamma_\infty\over 2})$ factors, these
equations are highly non-linear. However, using (\ref{eom}) it is
straightforward to check that $\gamma_\infty(t)$ is a {\it constant of
motion}. Hence, given any {\it one} solution, we can define a new time
coordinate $T$ on $M$ via a constant rescaling: $T := (\exp
{-{1\over 2}\gamma_\infty})\, t$. Then, the field $\psi$ satisfies the
following {\it linear} second-order equation of motion:
\be
   -\frac{\partial^2 \psi}{ {\partial T}^2} + 
   \frac{\partial^2 \psi}{{\partial R}^2} + 
   \frac{1}{R}\frac{\partial \psi}{\partial R} = 0.
\label{17a}
\ee 
This is {\it exactly} the Klein-Gordon equation for a scalar field
propagating on a Minkowskian background $g^o_{ab}$,
given by: 
\be 
g^o_{ab} = - \nabla_a T \nabla_b T + \nabla_a R
\nabla_b R + R^2 \nabla_a \sigma \nabla_b \sigma\, .
\label{17c}
\ee 
Thus, a remarkable simplification has occurred. We can just solve for
a free scalar field $\psi$ in Minkowski space $(M, g^o_{ab})$, {\it
define} a function $\gamma$ through (\ref{15}), and construct a curved
metric $g_{ab}$ through (\ref{17}).  Then the pair $(g_{ab}, \psi)$
satisfies the non-linear Einstein-Klein-Gordon equations.

This decoupling is not surprising from the space-time
perspective. Indeed, it has been exploited repeatedly in the
literature. However, it is illuminating to see how the decoupling
comes about from a phase space perspective especially since the
dynamics of the true degrees of freedom is driven only by the boundary
term Hamiltonian which, furthermore, seems quite complicated at
first sight. Note also that, while from a space-time perspective the
passage between $t$ and $T$ involves only a constant rescaling, since
the constant varies from solution to solution, from a phase space
perspective it is a rather complicated, ``q-number''
transformation. Thus, in quantum theory, if one variable in the pair
$(t, T)$ is taken as a ``time-parameter'', the other will be a genuine
operator. It is therefore instructive to contrast the two notions of
time. By construction, $t$ can be identified with the affine parameter
along the Hamiltonian vector field defined by (\ref{ham}) on the phase
space. Given any dynamical trajectory, we obtain a space-time metric
$g_{ab}$ and $t$ can then be thought of a time coordinate on $M$ with
the property that $\partial/\partial t$ generates an {\it unit} time
translation at infinity. The parameter $T$, on the other hand does not
have a direct and simple physical interpretation in our phase space
framework. Its most direct interpretation comes from the fiducial
Minkowskian metric $g^o_{ab}$ on $M$. Even at
infinity, the norm of the vector field $\partial/\partial T$ varies
from one physical metric $g_{ab}$ to another. For the decoupling
procedure, on the other hand, it is natural to fix, once and for all,
the Minkowskian metric $g^o_{ab}$ on $M$ and regard
$g_{ab}$ simply as a ``derived'' quantity. Then $T$ does have a
natural interpretation of time. Finally, note that this somewhat
peculiar situation arose because, in 2+1 dimensions, the physical
metrics $g_{ab}$ do not approach a fixed Minkowskian metric even at
infinity (or alternatively, because in 3+1 dimensions, cylindrical
waves fail to be asymptotically flat in the conventional sense.)

We will conclude this section with a remark. To begin with, one can
ignore the broad physical problem of interest and focus just on a free
scalar field satisfying the wave equation on the Minkowskian
background $(M, {g}^o_{ab})$. The phase space for this system is the
same as our reduced phase space and the Hamiltonian is given by
$\gamma_\infty$. However, $\gamma_\infty$ does {\it not} have a direct
physical interpretation in terms of the original, coupled system; the
physical energy of our system is given by (\ref{energy}).

\section{Holonomy }
\label{s3}

As explained in the introduction, there is a non-perturbative approach
\cite{aa:91} approach to quantum general relativity in 3+1 dimensions
which is based on the assumption that traces of holonomies of a
certain connection are well-defined operators in the quantum theory.
We would like to investigate the status of this assumption in the
context of our midi-superspace. Therefore, in this section, we will
make a short detour to compute the holonomy in question in the
classical theory. Readers who are not familiar with this approach to
quantum gravity may skip this section without loss of continuity.

In the first-order (Palatini) formalism for $2+1$ general relativity
the fundamental variables are triads $e_a^I$ and connection 1-forms
which take values in the Lie-algebra of $SU(1,1)$ \cite{2+1, aa:91}.
Let us denote the $SO(2,1)$ connection by ${}^3\!A_a^I$ and its
pull-back to the 2-dimensional slice $\Sigma$ by $A_a^I$, where
$I,J,\cdots =0,1,2$ are internal indices with respect to a basis
$\tau_I$ in the Lie algebra of $SU(1,1)$.  The internal indices are
raised and lowered with a Minkowski metric $\eta_{IJ}$ with signature
$(-,+,+)$.

\subsection{$SO(2,1)$ Connection}\label{s3.1}

To obtain the internal connection for the space-time metric
(\ref{17}), we need to fix the internal (i.e., $SU(1,1)$) gauge.  This
is accomplished by fixing the triads $e_a^I$.  Our choice will be:

\be
 e_a^I \tau_I =  {\sqrt 2} e^{{1\over 2}(\gamma(R,t)-\gamma_\infty)}
		  (\nabla_a t) \, \tau_0 + 
		 {\sqrt 2} e^{{1\over 2}\gamma(R,t)} (\nabla_a R) \, 
		 \tau_1 + {\sqrt 2} R  (\nabla_a \sigma) \, \tau_2 
\label{19a}                  
\ee

\ni
It is straightforward to check that the space-time  metric (\ref{17}) 
is recovered via $g_{ab} = \eta_{IJ} e_a^I e_b^J$ with the convention 
$\eta_{IJ} = 2 \, Tr (\tau_I \tau_J )$.

The triad determines the (Christoffel symbols and the) internal
connection ${}^3\!A_a^I$ uniquely.  Its pull-back to the spatial slice
$\Sigma$ turns out to be:
\be
A_a = A_a^I \, \tau_I = \frac{\dot\gamma}{2}\, e^{{1\over 2}
{\gamma_\infty}}\, (\nabla_a R) \, \tau_2 + e^{-{1\over 2}\gamma} 
(\nabla_a \sigma) \, \tau_0.
\label{20}
\ee
Note, however, that since $R, \sigma$ fail to be smooth at $R=0$
our connection also fails to be smooth there.  However, our boundary
conditions do ensure that all physical fields are smooth at the origin.
Thus, this singularity is merely a reflection of a bad choice of gauge
(which has in effect introduced a ``source'' at the origin).  We can
remedy this situation by a gauge transformation. The general form
of gauge transformations is: 
\be
{A'}_a = g A_a g^{-1} - (\partial_a g) g^{-1} \quad{\rm with}\quad
       g=e^{\tau_I \Lambda^I (R,\sigma)}.
\label{21}
\ee 
By choosing the transformation parameters to be $\Lambda^0 =
e^{-{1\over 2}\gamma(0)} \sigma$ and $\Lambda^1 = \Lambda^2 = 0$, we
obtain a smooth connection as desired:
\be
{A'}_a = {A'}_a^I \, \tau_I = \frac{\dot\gamma}{2} 
e^{{1\over 2}\gamma_\infty} \nabla_a R \,
[ \cos \sigma \, \tau_2 - \sin \sigma \, \tau_1 ] + 
[e^{-{1\over 2}\gamma} - 1] \nabla_a \sigma \, \tau_0.
\label{22}
\ee

\subsection{Holonomy computation}\label{s3.2}

The holonomy  of ${A'}_a^I$  along a loop $\eta$ is given by a path 
ordered exponential of the integral of ${A'}_a^I$ along $\eta$:
\be 
U_\eta[A] := {\cal P} \exp \left(\oint _\eta A_a dS^a\right).
\label{23}
\ee
For quantum considerations, it turns out that the most interesting
loops are the integral curves of the rotational Killing vector
$\sigma^a$. Note that, along these curves, only the second term in the
expression (\ref{22}) of the connection contributes. Since the
internal vector in this term is constant, this part of the connection
is effectively Abelian.  Recall that in the case of an Abelian
connection the path ordered exponential reduces to an ordinary
exponential.  Hence, if $\eta$ is chosen to be the integral curve of
the Killing field with radius $r_o$, the holonomy can be easily
evaluated. We have:
\be
U_\eta[A'] = \cos \left[ \pi \left( 1 - e^{-{1\over 2}\gamma(r_o)} 
\right)\right] - 2 \tau_0 \sin \left[ \pi \left( 1 - e^{-{1\over 2}
\gamma(r_o)}\right)\right]
\label{24}
\ee
where we have used the fundamental representation of $SU(1,1)$. For our
purposes, it will suffice to consider these particular loops. 

Of special interest to the quantization program under consideration
are the functions $T^0_\eta[A]$ of connections defined by the trace of
the holonomy. Taking the trace of (\ref{24}) yields
\be
T^0_\eta[A'] = 2 \cos \left[ \pi 
\left( 1 - e^{-{1\over 2}\gamma(r_o)}\right)\right]. 
\label{25}
\ee
Note, incidentally, that if $\eta$ is chosen to be the loop at
infinity, $T^0_\eta[A]$ reduces to a simple function of the total
energy of the coupled system. For the reduced system, $\gamma(r_o)$
represents precisely the energy of $\psi$ in a box of radius $r_o$
(where $\psi$ is regarded as a scalar field propagating on the
Minkowskian background.) The question of whether the $T^0_\eta$ can be
promoted to a well-defined operator will therefore reduce to the
question of whether the operator corresponding the energy of a scalar
field in a box can be satisfactorily regulated.

\section{Quantum Theory}
\label{s4}

\subsection{Quantization}\label{s4.1}

The reduced phase space of section \ref{s2.4} serves as the natural
point of departure for quantization. Since the constraints have been
solved, the algebra ${\cal A}$ of observables is easy to
construct. The obvious complete set of classical observables is given
by the smeared fields and momenta, $\psi(f) := \int dr f(r)\psi(r)$
and $p_\psi(g) := \int dr g(r)p_\psi(r)$, where $f, g$ belong to the
Schwartz space ${\cal S}$ of smooth test functions with rapid decay at
infinity.  Thus, the quantum algebra ${\cal A}$ is generated by
operators $\hat\psi(f)$ and $\hat{p}_\psi(g)$, subject to the
canonical commutation relations:
\be 
[{\hat \psi} (f),{\hat \psi} (g)]=0, \;\;\;[{\hat p}_\psi (f),
{\hat p}_\psi (g)]=0, \;\;\; [ {\hat \psi} (f), {\hat p}_\psi (g)]
= i \int dr fg\,\, \hat{I}.
\label{ccr}
\ee
Our task is to find a representation of ${\cal A}$ which,
furthermore, carries a well-defined Hamiltonian operator $\hat{H}$,
the quantum analog of   ${\textstyle{1\over 4G}}(1 - \exp
(- {\frac{\gamma_\infty}{2}}) )$.

For technical simplicity, we will regard $\hat{\psi}$ and
$\hat{p}_\psi$ as operator-valued distributions in two (space)
dimensions and incorporate rotational symmetry by restricting the
states to be axi-symmetric at the very end. Our experience from low
dimensional, interacting scalar quantum field theories now suggests
that we use as our Hilbert space {\cal H}= $L^2({\cal S}', d\mu)$
where ${\cal S}'$ is the space of all tempered distributions on $R^2$,
and $\mu$ a suitable measure thereon. (For details, see, e.g.,
\cite{gj}). Since $\gamma_\infty$ is the Hamiltonian of the free
scalar field in Minkowski space, to make the quantum Hamiltonian
operator well-defined, it is natural to use for $\mu$ the standard
Gaussian measure for a free, massless scalar field with covariance
$\textstyle{1\over 2}\triangle^{-{1\over 2}}$, where $\triangle$ is
the Laplacian on $R^2$ with respect to the flat metric\\
\[ q^o_{ab} = \nabla_a R \nabla_b R 
+ r^2 \nabla_a\theta \nabla_b \theta .\]
Thus, $\mu$ is defined by
\be \int_{{\cal S}'} d\mu \,\, e^{i\int d^2x f(\vec{x})
\tilde\psi(\vec{x})}\,\, = \, e^{-{1\over 2}\int d^2x f(\vec{x}) 
\triangle^{-{1\over 2}} f(\vec{x})}\, , \ee
where $\tilde\psi \in {\cal S}'$. (Heuristically, ``$d\mu = [\exp
{-{\textstyle {1\over 2}}}\int d^2x (\psi \triangle^{\textstyle
{{1\over 2}}}\psi)]{\cal D}\psi$''.) The action of the basic operators
is then given by:
\be
{\hat \psi}(f)\cdot\Psi(\psi) = \left(\int d^2 x f \psi \right) 
\Psi(\psi) \quad{\rm and}\quad
{\hat p}_\psi(g) \cdot \Psi(\psi)= -i\int d^2x\, [g\, 
\frac{\delta}{\delta \psi} + \frac{1}{2}\psi 
\triangle^{1\over 2} g]\, \Psi(\psi) 
\label{26c}
\ee
where $\Psi$ belongs to the dense sub-space of cylindrical functions in
${\cal H}$. The operators $\hat\psi(f)$ and $\hat{p}_\psi (f)$ admit
self-adjoint extensions to ${\cal H}$. We will see below that the
Hamiltonian is also represented by a self-adjoint operator and that,
like its classical counterpart, it is positive.

This choice of representation is also suggested by the mathematical
equivalence between our physical system and a free massless scalar
field on Minkowski space defined by $g^o_{ab}$ (see Eq
(\ref{17a})). Thus, although our viewpoint is somewhat different, our
final choice of representation is the same as that of Refs
\cite{kk71,ma87}.

In a more familiar terminology, our representation can be obtained by
introducing an operator-valued distribution $\hat{\psi}(\vec{x}, T)$ in the
fictitious Minkowskian background $(M, g^o_{ab})$:
\be
{\hat \psi}({\vec x},T)    =  \frac{1}{2\pi} \int 
\frac{d^2 k}{\sqrt{2\omega_k}} \left[ {\hat A}({\vec k}) \right. 
e^{i({\vec k}\cdot{\vec x}-\omega_k T)}+ {\hat A}^\dagger 
({\vec k})  \left. e^{-i({\vec k}\cdot{\vec x}-\omega_k T)} 
\right], 
\label{29a}
\ee
where $\omega_k = \sqrt{{\vec k}\cdot {\vec k}}$, and $\hat{A}
(\vec{k})$ and $\hat{A}^\dagger(\vec{k})$ are the standard creation
and annihilation operators. The Hilbert space ${\cal H}$ can be
generated by repeated actions of creation operators on the vacuum.
There is a well-defined self-adjoint operator $\hat{L}_\sigma$ on
${\cal H}$ which represents the total angular momentum along the
Killing field $\partial/\partial\sigma$. The physical Hilbert space
${\cal H}_P$ is the eigenspace of $\hat{L}_\sigma$ with zero
eigenvalue. Since zero is a discrete eigenvalue, ${\cal H}_P$ is a
sub-space of ${\cal H}$.

The physical Hilbert space can also be obtained more directly by
using, instead of (\ref{29a}), an operator valued distribution in
which the zero angular momentum constraint has already been
incorporated, namely,
%Monica: Is there a factor of k in the integral?
\be
{\hat \psi}(R,T) = \int_0^\infty d k \left[ f_k^+ (R,T) {\hat A}(k) 
+ \right. \left. f_k^- (R,T) {\hat A}^\dagger (k)\right]\, .
\label{29b}
\ee
Here $f_k^+(R,T) = [f_k^- (R,T)]^*=\frac{1}{\sqrt 2} J_0(kR)
e^{-i\omega_k T}$, where, from now on, $J_n(kR)$ will denote the $n$-th
order Bessel function of the first kind.  Note that $f^+_k(R)$ are
solutions of the equation of motion (\ref{17a}) and provide an
orthonormal basis for the one-particle Hilbert space with respect to
the Klein-Gordon inner-product. (Our normalization is such that the
creation and annihilation operators satisfy the commutation relations
$[{\hat A}(k), {\hat A}^\dagger(k')] = \delta(k,k')$.) The physical
Hilbert space ${\cal H}_P$ can be generated by repeatedly acting on
the vacuum by the creation operators $\hat{A}^\dagger(k)$. In what
follows, we will use both the two dimensional as well as the one
dimensional descriptions given by (\ref{29a}) and (\ref{29b}).

We  will  conclude this sub-section with three remarks.
\begin{itemize}
\item[1)] Since the physical Hilbert space has a Fock structure, it is
tempting to refer to the quanta created by $\hat{A}^\dagger(k)$ as
(scalar) ``particles'' and we will often do so. Note, however, that from
the point of view of the coupled Einstein-Klein-Gordon system we began
with, this description is gauge dependent. The system has one local
degree of freedom and we chose to put it in the scalar field. Another
gauge choice could put it in the gravitational field and the
interpretation of quantum states would then be different.  However,
the interpretation {\it is} unambiguous at null infinity  --i.e, for
asymptotic states--  because one does not need to fix gauge there
(see below).
\item[2)] We now have the full Hilbert space of states. So, it is
natural to examine if one can generate a picture of space-{\it time}
--as opposed to just spatial-- quantum geometry in spite of our use of
the canonical approach. As one might expect from our gauge-fixing
procedure, the answer is in the affirmative. In the fixed chart $(T, R,
\sigma)$ on $M$, the metric operator can be (heuristically) written
as:
\be
``\;{\hat g}_{ab}\, =\,\, :\! e^{{\hat \gamma} (R,T)}: 
	  (- \nabla_a T \nabla_b T + \nabla_a R \nabla_b R )
	  + R^2 \nabla_a \sigma \nabla_b \sigma\; {\hbox{''}}, 
\label{mo}
\ee 
where, as usual, the double-dots indicate normal ordering. (The reason
behind the qualification ``heuristic'' and the quotes will become
clear in section \ref{s4.3}.) 

We can now ask if there are well-defined semi-classical states peaked
at classical solutions. The answer is again in the
affirmative. Consider, in the Fock space, a coherent state $|\Psi_c\!
>$ which is peaked at a classical solution $c(R,T)$ of (\ref{17a}). In
the configuration representation, these are Gaussians for which the
uncertainty in the field operator and its momentum are ``shared
equally'', the product of the two uncertainties being minimum {\it for
all times} $T$. On these states, the expectation value of the metric
operator (\ref{mo}) {\it is well-defined} and is given just by
\be
<\Psi_c|{\hat g}_{ab}|\Psi_c> = e^{ \gamma[c, p_c] } (- \nabla_a T 
\nabla_b T + \nabla_a R \nabla_b R ) + R^2 \nabla_a \sigma 
\nabla_b \sigma,
\label{32}
\ee
where $\gamma[c, p_c]$ is the right side of (\ref{16}), evaluated on
the initial data of the classical solution $c$. Thus, every coherent
states in our physical Hilbert space ${\cal H}_o$ remains peaked at a
classical scalar field $c$ {\it and} a metric $g_{ab}$, satisfying the
coupled Einstein-Klein-Gordon equation. While the result is
technically rather simple, conceptually it is somewhat
surprising. For, the coupled system satisfies highly non-linear
equations and the wave packets do not disperse in spite of these
non-linearities.
\item[3)] It is well-known that there exist an infinite number of
unitarily inequivalent representations of the algebra ${\cal A}$. Our
additional requirements are that the Hamiltonian operator be
well-defined and that the physical states be invariant under the
rotational symmetry corresponding to $\partial/\partial\sigma$. 
Unfortunately, these requirements by themselves are not strong enough
to select a representation uniquely. To single out the Fock
representation in Minkowskian quantum field theories, one needs
additional conditions that refer to the action of the Poincar\'e
group. In our case, the Minkowski space-time $(M, g^o_{ab})$ is only a
fictitious background and its Poincar\'e group has no physical
significance in the full, coupled system.

Nonetheless, it {\it is} possible to single out our representation by
two methods. The first involves the imposition of reality conditions
as indicated in \cite{aa:91}: The measure $\mu$ on ${\cal S}'$ is
singled out by the condition that the operators $\hat\psi (f)$ and 
$\hat{p}_\psi(g)$ of (\ref{26c}) be self-adjoint. The second method
invokes the S-matrix theory. It turns out that the Einstein-Rosen
waves are all asymptotically flat at null infinity in 2+1 dimensions
\cite{aajbbs}. Furthermore, the classical $S$-matrix is well-defined:
the data on past null infinity determines the solution uniquely which
in turn determines the data on future null infinity. Hence, it is
natural to use the asymptotic quantization scheme \cite{aa:aq} to
quantize the coupled system {\it at} past and future null infinity. It
turns out that our Fock representation is naturally isomorphic to the
simplest representation obtained by asymptotic quantization (either at
past {\it or} future null infinity). Details will appear elsewhere.
\end{itemize}

\subsection{Hamiltonian and Time}\label{s4.2}

Recall that, after reduction, the classical Hamiltonian is given by
$H={\textstyle{1\over 4G}}[1 - \exp -({\frac{1}{2}}\gamma_\infty)]$.
Since $\gamma_\infty$ is the Hamiltonian of a free scalar field in
Minkowski space, the normal-ordered operator $:\!\hat{\gamma}_\infty
\! :$ admits the standard self-adjoint extension which, for
simplicity, we will denote also by $:\!  \hat{\gamma}_\infty\!:$. 
Then, the standard spectral theorems ensure that
\be\label{H}
\hat{H} \, :=\, {1\over 4G}(1 - e^{ - {1\over 2} :\! 
\hat{\gamma}_\infty\! :} ) \,\equiv\, {1\over 4G}(1 - 
e^{-\int kdk \hat{A}^\dagger (k)\hat{A}(k)})
\ee
is a well-defined, self-adjoint operator. Since $:\!
\hat{\gamma}_\infty \! :$ is a non-negative, unbounded operator and
since $f(\lambda)= (1- e^{-{\lambda\over 2}})$ takes values in $[0, 1]$
for $\lambda \in [0, \infty]$, it follows that the spectrum of $H$ is
given by $[0, 1/4G]$. If we consider states in ${\cal H}_P$ with
higher and higher frequency, the expectation value of
$\hat{\gamma}_\infty$ --i.e., the energy in the field from the
mathematical, Minkowskian perspective-- increases unboundedly.
However, the expectation value of the {\it physical} Hamiltonian
$\hat{H}$ remains bounded and tends to the limit $1/4G$. Thus, the
situation is completely analogous to that in the classical theory
\cite{aamv94}.

Let us now examine the ground state. Since $|0\! >$ is the unique
ground state of $:\!\hat{\gamma}_\infty\! :$ on ${\cal H}_P$, it
follows immediately that it is also the unique ground state of
$\hat{H}$.  Since $|0\! >$ is, in particular, a coherent state, it is
peaked at a classical solution to the coupled system. As one might
expect, the solution is: $\psi = 0$ and $g_{ab} = {g}^o_{ab}$. Thus,
the quantum ground state is peaked on Minkowski space-time. The ground
state geometry is thus quite tame, there is no evidence of wild
fluctuations at the Planck scale.

What is the situation with general coherent states? Given a coherent
state $|\Psi_c\! > := \exp [\int dk c(k) \hat{A}^\dagger(k)]\cdot
|0\! >$, peaked at a classical solution $c$, we have:
\be
[\exp\, -{\textstyle {1\over 2}}:\! \hat{\gamma}_\infty\! :]\,
\cdot \Psi_c = [\exp\, \int dk e^k c(k) \hat{A}^\dagger(k)]\cdot 
|0\! > =: \, \Psi_{c'}
\ee
where, $c'(k) = e^k c(k)$. Thus, the image is again a coherent state
but its peak is shifted. Therefore, the expectation value of
the Hamiltonian in a coherent state $\Psi_{c}$ is given by:
\be \label{qe}
{{<\! \Psi_c\, ,\,\hat{H}\cdot \Psi_c\!>}\over {<\!\Psi_c\, ,\,
\Psi_c\!>}}\, =\, {1\over 4G}\,[ 1 - \exp\, \textstyle{{1\over \hbar}}
\int dk (e^{-\hbar k}-1) |c(k)|^2\, ]\, , 
\ee
where, to bring out the quantum effects, we have restored the factors
of $\hbar$. (Recall also that, from the perspective of the
2+1-dimensional theory, the scalar field has to be rescaled by factors
involving $\sqrt{G}$. The net effect is to replace $\hbar$ in (\ref
{qe}) by $\hbar G$ which has the physical dimension of length.)  By
contrast, the classical energy (\ref{energy}) of the solution to the
Einstein-Klein-Gordon equation determined by $c$ is $E(c) =
{\textstyle{1\over 4G}}[1 - \exp\, -\int dk k |c(k)|^2]$. If we expand
out $\exp \hbar k$ in (\ref{qe}), the leading term yields the
classical answer. In general, the classical energy is a good
approximation to the expectation value of the quantum Hamiltonian if
$c(k)$ is concentrated on low frequencies.  Quantum corrections (of
order $(G\hbar)$ and higher) become more and more significant if the
support of $c(k)$ is shifted to higher and higher frequencies.

Next, let us consider the issue of time. Recall that, in the classical
theory, the Hamiltonian evolution is tied to time $t$, the affine
parameter along the Hamiltonian vector field in the phase space. Each
dynamical trajectory gives rise to a space-time and $t$ can then be
interpreted as a time coordinate in {\it that} space-time, $\partial/
\partial t$ being an unit asymptotic time translation. {}From the
decoupling viewpoint, on the other hand, it is the variable $T$ that
arises naturally; it represents time in the fixed Minkowskian
background. What is the situation in the quantum theory? Now, our
measure $\mu$ on ${\cal S}'$ which dictates the Hilbert structure is
rooted in the flat 2-geometry induced by $g^o_{ab}$ or, alternatively,
in the positive and negative frequency decomposition with respect to
the Minkowskian time $T$. Indeed, since the field equation (\ref{eom})
in terms of $t$ is non-linear, positive frequency decomposition with
respect to $t$ is not meaningful apriori. Thus, while $t$ and $T$ are
on equal footing in the classical theory, our choice of representation
breaks this symmetry in the quantum theory.

We can mimic the situation in the classical theory and introduce
a dynamical parameter $\lambda$  --analogous to the classical $t$--  
associated with the Hamiltonian:
\be\label{se}
i\hbar \frac{\partial \Psi}{\partial \lambda} = \hat{H}\cdot \Psi \, . 
\ee
However, unlike in the classical theory, now a solution to the
dynamical equation does {\it not} define a hyperbolic space-time and
hence we can not interpret $\lambda$ as a time parameter in the
familiar sense, i.e., in space-time terms. However, a key
simplification occurs if we restrict ourselves to coherent states
$\Psi_c$ . Since each of these states is peaked at a classical
space-time, we can ask if, given any one of these states, we can
interpret $\lambda$ as a time parameter in the corresponding classical
space-time. The answer is in the affirmative.  In fact $\lambda$ can
be identified with the time coordinate $t$ of that space-time! Thus,
as one might have hoped, the familiar notion of time re-emerges in the
semi-classical regime. In the full quantum theory, however, the
dynamical parameter defined by the Hamiltonian does not have a simple
space-time interpretation.

We will conclude this discussion with a remark. There is an obvious
alternative form for the Hamiltonian: We can further normal-order
$\hat{H}$ and define a new Hamiltonian $\hat{H}' =\, :\! \hat{H}\! :$.
One can verify that $\hat{H}'$ is densely defined and admits a
self-adjoint extension. It also annihilates $|0\! >$. Furthermore, the
expectation values of $\hat{H}'$ on a coherent  state $|\Psi_c\!>$
equals the classical energy of $c$. It thus appears to be an
attractive alternative. However, its spectrum is the {\it entire real
line}!  This comes about because the overall normal ordering ensures
that, while acting on $n$-particle states, only the first $n+1$ terms
in the expansion of the exponential in $\hat{H}'$ have non-vanishing
contributions. Thus, for example, on $1$-particle states, $\hat{H}'$
has the same action as $\textstyle{1\over 8}\,:\!\hat\gamma_\infty\!
:$ which is unbounded above. Similarly, on two particle states, it is
unbounded below. Given that the classically allowed energy values lie
in the interval $[0,1/4G]$, we can not take ${\hat{H}}'$ as the
physically admissible quantum analog of the classical Hamiltonian.

\subsection{Metric operator}\label{s4.3}

Since we are dealing with a system with an infinite number of degrees
of freedom, operators corresponding to physical observables have to be
regulated. For the Hamiltonian, this was achieved via normal ordering.
In this section, we will focus on the metric operator.

A formal expression for the metric operator was already given in
(\ref{mo}), where regularization again consisted of normal ordering.
Consider the sub-space of ${\cal H_P}$ which is spanned by finite
linear combinations of coherent states. It is easy to show that the
sub-space is dense and that the matrix elements of the metric operator
$\hat{g}_{ab}$ are well-defined on it. Thus, the formal expression
(\ref{mo}) does lead to a well-defined {\it quadratic form}; in a
field theory terminology, $\hat{g}_{ab}$ exists in the LSZ sense.
However, this does {\it not} imply that $\hat{g}_{ab}$ is well-defined
{\it as an operator} on this sub-space. Note that this is {\it not} a
peculiarity of quantum field theory; one encounters such situations
already in non-relativistic quantum mechanics. Consider, for example,
a $1$-dimensional harmonic oscillator. The operator $\exp (\alpha
a^\dagger a^\dagger)$ has finite matrix elements on the basis $|n\! >$
for all complex numbers $\alpha$. However, if $|\alpha|\! >1$, the
norm $||e^{\alpha a^{\dagger}a^{\dagger}}|n\! >||$ diverges for any
$|n\!>$, whence the operator fails to be defined on the sub-space
spanned by these basis vectors.

It turns out that the situation with the metric operator is quite
analogous (which is the reason behind the quotes in (\ref{mo})). To see
this, let us begin with the first non-trivial term in the expansion of
$\hat{g}_{RR}$ or $\hat{g}_{TT}$. Setting for simplicity $T = 0$ in
(\ref{29b}), we have:
\ba 
:\! {\hat \gamma}(R)\! : \, =  \frac{1}{2} \int d k_1 \int d k_2 &&
\left[ 2F_{+}(R,k_1,k_2)\left( {\hat A}^\dagger (k_1){\hat A}(k_2) 
\right)\right.  \nonumber\\
&& + \, F_{-}(R,k_1,k_2) \left( {\hat A}(k_1){\hat A}(k_2) 
\right.  + \left. \left. {\hat A}^\dagger (k_1) 
{\hat A}^\dagger (k_2) 
\right)\right],
\label{30} 
\ea
where
\be 
F_{\pm}(R,k_1,k_2)= \pm k_1k_2\, \int_0^R r dr\, \left(J_0 (k_1 r)
J_0 (k_2 r) \pm \right. \left. J_1 (k_1 r) J_1 (k_2 r) \right).
\label{31}
\ee
For any fixed $R$, one can regard the coefficient $F_{-}(R, k_1, k_2)$
of $\hat{A}^\dagger(k_1)\hat{A}^\dagger(k_2)$ as a ``potential
2-particle state'' in the Fock space. However, a direct calculation
shows that its norm is ultra-violet divergent. This immediately
implies that the norm \\ $||:\! \hat\gamma(R)\! :|0\! >||$ also
diverges, whence the operator fails to be well-defined on the vacuum
state. Further calculations show that the same result holds for any
coherent state.

What is the origin of this divergence? Recall that $:\! {\hat
\gamma}(R,T)\! :$, obtained by promoting (\ref{15}) to an operator,
has the same functional form as the restriction of the Hamiltonian of
a scalar field to a box of size $R$. That is,
\be \label{31aa}
:\! {\hat \gamma}(R)\! :\,  =  \frac{1}{2} \int_0^\infty dr
\theta (R-r)\, :\! ( \frac{{{\hat p}_\psi}^2}{r} 
+ r {({\hat \psi}'})^2 )\! :, 
\ee
where $\theta (R-r)$ denotes the Heaviside step-function, which equals
$1$ if $r<R$ and $0$ otherwise. Normal ordering softens the
singularity that arises from the fact that fields are being multiplied
at the same point. However, this turns out to be insufficient because
of two simultaneous pathologies: the operator contains products of
derivatives of the field ${\hat \psi}(R,T)$ and these are integrated
on a region with {\it sharp} boundary.

Now, a natural strategy to obtain a well-defined metric operator in
such circumstance is to soften the sharp boundary of the box.  This
can be achieved by replacing the Heaviside function $\theta (R-r)$ in
(\ref{31aa}) with a smooth function $f_R(r)$ which equals $1$ for $r
\leq R- \epsilon$, then it smoothly decreases to zero and equals zero
for $r \geq R+ \epsilon$, where $\epsilon$ is a small parameter. 
An example of such a regulator is:
\[
f_R(r) = \cases{ 1,  & if $r \le R-\epsilon$, \cr
		 \exp \left(-{{4\epsilon^2}\over 
		 {[r- (R+\epsilon)]^2}}+1\right), & if $R-\epsilon
		 \le r \le R+\epsilon$, \cr
		 0, & if $r\ge R+\epsilon$.\cr}
\]
Now, in Minkowskian field theories, while one can begin with such a
regulator, after suitable renormalization, one has to take the
regulator away to ensure Poincar\'e invariance. In the present case,
however, we need only respect the rotational symmetry and hence there
is no apriori need to take the limit $\epsilon \to 0$. Indeed, the
Planck length is now a natural candidate for $\epsilon$.

Let us therefore fix a regulator $f_R$ and consider
the smeared version of (\ref{30}):
\ba
:\! {\hat \gamma}(f_R, T)\! :\,\, =  \frac{1}{2} \int d k_1 \int d k_2 
& & \left[2 F_{+}(f_R,k_1,k_2)\left( {\hat A}^\dagger (k_1){\hat A}
(k_2)e^{i(k_1 - k_2)T}\right)\right. \nonumber \\
& & +\,  F_{-}(f_R,k_1,k_2) 
 \left( {\hat A}(k_1){\hat A}(k_2) e^{-i(k_1 + k_2)T}\right.\nonumber\\
& & + \left. \left. {\hat A}^\dagger (k_1) {\hat A}^\dagger (k_2) 
e^{i(k_1 + k_2)T} \right)\right]\, 
\label{31a} 
\ea
where,
\be
F_{\pm}(f_R,k_1,k_2)= \pm k_1 k_2\, \int_0^\infty f_R(r) r 
\left(J_0 (k_1 r)  J_0 (k_2 r) \pm \right. \left. 
J_1 (k_1 r) J_1 (k_2 r) \right).
\label{31b}
\ee
The rest of this section is devoted to showing that this operator is
well-defined so long as the smearing function $f_R$ belongs to the
Schwartz space ${\cal S}$.

The proof is technically simpler if we adopt the $2$-dimensional
version of the Fock space introduced before (see (\ref{29a})).  For,
we can then mimic the proofs of analogous statements from \cite{cj70}.
Now, we can take as our smearing fields, elements $f_R({\vec x})$ of
the Schwartz space on ${R}^2$.  (Thus, the results will in fact be
slightly more general than what is need; $f_R(r)$ above is a special
case of $f_R({\vec x})$.)

Let us then write the smeared version of the operator (\ref{31aa})
expressed in terms of the creation and annihilation operators given by
(\ref{29a}). We have:
\ba \label{33} 
:\! {\hat \gamma}(f_R,T)\! :\, = \frac{1}{8\pi} \int d^2 k_1
\int d^2 k_2 & & \left[2 G_{+}(f_R,{\vec k}_1 , {\vec k}_2) \right.  
{\hat A}^\dagger ({\vec k}_1){\hat A}({\vec k}_2)
e^{i(\omega_{k_1}-\omega_{k_2})T} \nonumber \\ 
& & -\, G_{-}(f_R,{\vec k}_1 ,{\vec k}_2)\,
\left( {\hat A}({\vec k}_1) {\hat A}({\vec k}_2) \right.  
e^{-i(\omega_{k_1}+\omega_{k_2})T}\nonumber\\
& & + \left. {\hat A}^\dagger ({\vec k}_1) {\hat A}^\dagger 
({\vec k}_2)\, e^{i(\omega_{k_1}+\omega_{k_2})T}
\right) \left. \right] 
\ea
where
\be \label{33a}
G_{\pm}(f_R,{\vec k}_1 , {\vec k}_2)= \pm \left( 
\frac{\omega_{k_1}\omega_{k_2} + {\vec k}_1 \cdot {\vec k}_2}
{\sqrt{\omega_{k_1} \omega_{k_2}}}\right) 
f({\vec k}_1 \mp {\vec k}_2 )\, ,
\ee
and  $f({\vec k}_1 \pm {\vec k}_2 )$ is the fourier transform of the 
smearing function,
\be
f({\vec k}_1 \pm {\vec k}_2 ) = \frac{1}{2\pi} \int d^2 x f_R({\vec x}) 
    e^{i ( {\vec k}_1 \pm {\vec k}_2 ) \cdot {\vec x}}.
\label{34}
\ee

Let us begin by showing that the action of the operator (\ref{33}) is
well-defined on the vacuum state. Since ${\hat A}({\vec k})$
annihilates the vacuum state, we have:

\be
||:\! {\hat \gamma}(f_R)\! :|0\! >||^2= \int d^2 k_1 \int d^2 k_2\,\,  
	 | G_{-}(f_R,{\vec k}_1 , {\vec k}_2)|^2\, .
\label{35}     
\ee
It follows immediately from (\ref{33a}) that this integral has no
infra-red divergences. Therefore, from now on, let us concentrate only
on the ultra-violet behavior of the integrand. The factor in the round
brackets is ultra-violet divergent. The multiplicative factor $f$
provides a damping, but only for large $|\vec{k_1} + \vec{k_2}|$.
However, using simple algebra one can bound $G_{-}(f_R,{\vec k}_1 ,
{\vec k}_2)$ of Eq (\ref{33a}) by
\be
|G_{-}(f_R,{\vec k}_1 , {\vec k}_2)| \leq \frac{|{\vec k}_1 
+{\vec k}_2 |^2 |f({\vec k}_1 + {\vec k}_2)}{\sqrt{\omega_{k_1} 
\omega_{k_2}}}.
\label{36}
\ee
Now, because the smearing function $f_R({\vec x})$ belongs to the
Schwartz space, its Fourier transform $f({\vec k}_1 + {\vec k}_2)$
falls faster than any polynomial in $|\vec{k_1}+\vec{k_2}|$. This in
turn implies that $G_{-}(f_R, \vec{k}_1, \vec{k}_2)$ is square
integrable.  Note that the smearing function plays a crucial role in
this argument. Had we replaced $f_R({\vec x})$ by the Heaviside
function $\theta$ the corresponding Fourier transformed function
$f({\vec k}_1 + {\vec k}_2)$ would behave as $1/|{\vec k}_1 +{\vec
k}_2|$ which would not be sufficient to ensure square-integrability of
$G_{-}(f_R,{\vec k}_1 , {\vec k}_2)$ (see (\ref{36})). Finally, as a
side remark, note that the procedure followed above to prove that
$G_{-}(f_R,{\vec k}_1 , {\vec k}_2)$ is square integrable does not go
through for $G_{+}$ because of the minus sign in the argument of the
function $f({\vec k}_1 - {\vec k}_2)$ (see (\ref{33a})).
 
Next, one can show that the action of this operator is in fact
well-defined on a generic n-particle state on the Fock space,
\be
|\Psi_n\! > \, = \, \int d^2k_1 \cdots d^2 k_n\,\, g^{(n)}
({\vec k}_1 , \cdots , 
{\vec k}_n) {\hat A}^\dagger ({\vec k}_1) \cdots 
{\hat A}^\dagger ({\vec k}_n)|0>,
\label{37}
\ee
where $g^{(n)}({\vec k}_1 , \cdots ,{\vec k}_n) = <{\vec
k}_1,\cdots,{\vec k}_n |\Psi_n\! >$, and $\int d^2 k
|g^{(n)}(\cdots,{\vec k},\cdots)|^2 < \infty$.  Now the terms
involving annihilation operators will also contribute. The final
result is that $||:\! {\hat \gamma}(f_R)\! :|\Psi_n\! >||$ is finite
provided that $|\Psi_n\! >$ is a state such that $\int d^2 k |{\vec
k}|^2 |g^{(n)}(\cdots,{\vec k},\cdots)|^2 < \infty$.  (This
restriction comes from the ``particle number preserving term'' in
(\ref{33}).)  Since finite linear combinations of these states form a
dense subset of the Hilbert space, we have now established that the
operator $:\! {\hat \gamma}(f_R)\! :$ is densely defined on ${\cal H_P}$.

By inspection, it also symmetric on this space. We will now show that
it admits a self-adjoint extension to ${\cal H}_P$. For this, by a
theorem due to Von-Neumann \cite{rs}, it is sufficient to exhibit on
${\cal H}_P$ an anti-linear operator $\hat{C}$ with $\hat{C}^2 =1$
which leaves the domain of $:\! {\hat \gamma}(f_R)\! :$ invariant and
commutes with it. We can take $\hat{C}$ to be the complex-conjugation
operator on ${\cal H}_P = L^2({\cal S}', d\mu)$. It is straightforward
to show that $\hat{C}$ commutes with $\hat\psi(\vec{x}, T)$ whence
$\hat{C} \hat{A}(\vec{k}) = \hat{A} (-\vec{k}) \hat{C}$, and, $\hat{C}
\hat{A}^\dagger(\vec{k}) = \hat{A}^\dagger(-\vec{k}) \hat{C}$.
Finally, since $G_{\pm} (f_R, \vec{k}_1, \vec{k}_2)$ is real and
equals $G_{\pm}(f_R, -\vec{k}_1, -\vec{k}_2)$, it follows that
$\hat{C}$ satisfies the conditions of Von-Neumann's theorem. Again,
for notational simplicity, we will denote the self-adjoint extension
also by $:\! {\hat \gamma}(f_R)\! :$.

We can now return to the metric. Since $:\! {\hat \gamma}(f_R)\! :$ is
a self-adjoint operator on ${\cal H}_P$, it follows that $\exp :\!
{\hat \gamma}(f_R)\! :$ is also self-adjoint. Thus, we can now give
meaning to the formal expression (\ref{mo}) and define a regulated
operator for the full space-{\it time} metric:
\be
{\hat g}_{ab}(f) =  e^{:{\hat \gamma:} (f_R,T)} 
(- \nabla_a T \nabla_b T + \nabla_a R \nabla_b R )
+ R^2 \nabla_a \sigma \nabla_b \sigma\, ,
\label{37ab}
\ee
{\it within canonical quantization}. In the classical theory, the
existence theorems ensure that a space-time metric can be recovered
from the canonical framework. There is, however, no such general
result in the quantum theory. Our success can be traced back to the
use of a well-suited gauge fixing procedure. (Whether a different
choice of gauge will give equivalent results is far from being clear.)

At first, it is somewhat confusing that while we do not need a
smearing function to obtain a well-defined quadratic form, we need one
to obtain a well-defined operator. Note however, that the situation is
rather similar even in the classical theory!  The metric component
$\exp \gamma(R)$ is a well-defined functional on (a dense sub-space of)
the reduced phase space. However, precisely because of the sharpness
of the boundary, this functional {\it fails} to give rise to a
well-defined Hamiltonian vector field. To obtain a Hamiltonian vector
field, one again needs to soften the boundaries using a smearing
function.  The fact that the unsmeared functional is well-defined is
analogous to the fact that, in the quantum theory, the quadratic form
is well-defined without smearing. The smeared quantum operator is the
analog of the smeared classical observable with a well-defined
Hamiltonian vector field. {}From this perspective, in fact it would have
been surprising if a self-adjoint metric operator had existed without
smearing; it would then have defined a 1-parameter group of motions on
the Hilbert space which would have no classical counterpart.

\subsection{Quantum geometry}\label{s4.4}

We will now briefly investigate three consequences of the results
obtained in the last three sub-sections. The discussion will be rather
general and we will only indicate the directions along which more
detailed work could be done.

The first concerns the issue of vacuum fluctuations of geometry. To
compute these, we need a well-defined operator; quadratic forms do not
suffice. Let us therefore consider the regulated metric operator
(\ref{37ab}). Since it is completely determined by $:\!\hat\gamma
(f_R, T)\! :$, let focus on this latter operator. The vacuum
expectation value of this operator is zero. However, because of the
vacuum fluctuations, there is a non-zero probability of finding other
values as well.  A qualitative measure of these probabilities is given
by the uncertainty:
\ba \label{37a}
\left[\delta :\! {\hat \gamma}(f_R,T)\! : \right]^2 \,\,  
&:=& \, <0|(:\! {\hat \gamma}(f_R,T)\! :)^2|0\! > -
<0|:\! {\hat \gamma}(f_R,T)\! :|0\! >^2\, \nonumber\\
&=& {\int dk_1 dk_2 |F_-(f_R,k_1,k_2)|^2}.
\ea

The right side is a measure of the fluctuation of the metric
coefficents around the mean.  An immediate consequence of the above
result is the existence of the fluctuations of the light cone. To see,
this, consider a vector $k^a$ in the tangent space of a point
$(T,R,\sigma)$ which is null with respect to $g^o_{ab}$. Now, due to
the vacuum fluctuations of the metric operator, the value of the norm
of $k^a$ is uncertain and, since the fluctuation can have either sign,
there is in general a non-zero probability for $k^a$ to be space-like
or time-like. The exception occurs if the vector $k^a$ is radial,
i.e., orthogonal to $\partial/\partial\sigma$. Then, because of the
specific form (\ref{37ab}) of the metric operator, $k^a$ continues to
be null. (Similar considerations obviously apply to time-like and
space-like vectors.) This simple example illustrates that, contrary to
an oft-expressed view, the canonical framework {\it is} capable of
addressing space-{\it time} issues such as the fluctuations of the
causal structure.

The second feature we wish to discuss concerns the commutator of the
metric operators at the same value of $T$. Again, in this calculation,
quadratic forms do not suffice and we must use the regulated operator
(\ref{37ab}). A straightforward calculation yields:
\ba \label{37aaa} 
[:\!  {\hat \gamma}(f_R): ,  :\! {\hat \gamma}(g_{R'}): ] =
\frac{i}{2} \int d^2x & &\left( f({\vec x})
\nabla^a g({\vec x}) - g({\vec x}) \nabla^a
f({\vec x})\right) \times \nonumber \\
& & :\! ( {\hat p}_\psi ({\vec x}) \nabla_a {\hat\psi}({\vec x})
+ \nabla_a {\hat\psi} ({\vec x}) {\hat p}_\psi ({\vec x})): \, .
\ea
Thus, the commutator does {\it not} vanish; the non-vanishing
contribution comes from the smeared boundary at the smaller of $R$ and
$R'$. At first the result seems surprising since $\hat\gamma(f_R)$ and
$\hat\gamma(g_{R'})$ dictate the ``value'' of the metric operator at
points $R$ and $R'$ which can be widely separated (and have the same
value of $T$). However, the result does not contradict any physical
principle. For, although the basic field operators $\hat\psi$ and
$\hat{p}_\psi$ associated with such points do commute, the metric
operator is a {\it non-local}  functional of these.

Indeed, the result has a classical analog. As we pointed out at the
end of the last sub-section, the unsmeared metric $g_{ab}$ does not
define a Hamiltonian vector field on the reduced phase space. Hence,
to evaluate Poisson brackets, we are forced to use the smeared
metric. Then, it is easy to verify that the Poisson brackets between
the functionals $\gamma(f_R)$ and $\gamma(g_{R'})$ fail to vanish even
when $R$ and $R'$ are widely separated. In fact these Poisson bracket
just mirror the commutators given above.
 
The last point we wish to discuss concerns the holonomies computed in
section \ref{s3}. We found that the expression of the holonomy
involves the exponential of the integral of the connection along a
loop on $\Sigma$. Now, as we indicated in the Introduction, there is a
canonical quantization program which is based on the assumption that
the quantum analogs of these holonomies are well-defined operators.
The present model provides a good testing ground for the validity of
this assumption.

To see that the issue is non-trivial, let us first recall the
situation in the well-understood Maxwell theory, say in 2+1
dimensions. There, the connection is generally promoted to an
operator-valued distribution and the holonomies (of real connections)
fail to be well-defined in the standard Fock representation. For, in a
2+1-dimensional theory, the operator-valued connection has to be
smeared with 2-dimensional test fields while loops have only
1-dimensional support. In the present case, we are also using a Fock
representation. A natural question therefore arises: Is the situation
then analogous to the Maxwell theory? If so, the basic assumption
mentioned above would fail to hold in our solution.

Now, because of axi-symmetry, smearing along a path in the radial
direction in effect corresponds to a 2-dimensional smearing. Hence,
the acid test is provided by loops $R= {\rm const}$ where one can not
take advantage of axi-symmetry. Can the classical expression
(\ref{25}) of the trace of the holonomy along such a loop, $\eta$,
be promoted to a well-defined, regulated operator? Following the
procedure we used in section \ref{s4.3}, we find that the answer is in
the affirmative. The quantum operator is given by:
\be
{\hat T}^0_\eta = 2 \cos 
\left[\pi \left( 1-e^{-{1\over 2}:\!\hat\gamma(f_R)\! :}\right)
\right],\label{38}
\ee
The standard spectral theorems ensure that the operator on the right
is well-defined, self-adjoint and has spectrum bounded between $-1$
and $+1$. Thus, the situation is very different from that in the
Maxwell case. Indeed, in the present case, it is the scalar field that
is subject to Fock quantization. The connection --like the metric-- is
a {\it non-local} functional of the elementary scalar field; its
expression involves 2-dimensional integrals of the basic fields. It is
because of this that the trace of the holonomy can be promoted to a
well-defined operator on ${\cal H}_P$. As in the case of the metric,
if we were interested only in quadratic forms, there would be no need
to use any smearing fields; they are needed only if one wishes to
obtain genuine operators.

\section{Discussion}\label{s5}

The mathematical structure of the classical Einstein-Rosen waves has
been well-known for a long time. In light of those results, it is not
at all surprising that the true degrees of freedom can be coded in a
scalar field satisfying the wave equation with respect to a fictitious
Minkowski space and quantization of this field in itself is trivial.
Thus, the underlying structure of our final theory is the expected
one.  The main purpose of the analysis was, rather, to apply the
standard canonical quantization method --which is applicable in the
more general context-- to arrive at this final picture
systematically. That is, since the model is technically sufficiently
simple to be exactly soluble, we used it to better understand the
standard quantization techniques and to probe conceptual and technical
issues of quantum general relativity.

Indeed, the analysis shed light on a number of these. At the classical
level, we saw that one can effectively exploit asymptotic flatness to
disentangle gauge from dynamics. Gauge conditions can be imposed to
handle constraints and to extract the true degrees of freedom. In the
final picture, we are still left with a non-trivial Hamiltonian.
Consequently, the issue of deparametrization never enters our
analysis. Similarly, we did not find it necessary to introduce ``clock
degrees of freedom'' \cite{jrct95} at infinity to extract dynamics.
In the quantum theory, we saw that there exist semi-classical states
which are peaked at classical solutions of the coupled Einstein-scalar
field system. The positive energy theorem goes over to the quantum
theory and the quantum Hamiltonian has the same upper bound as the
classical one. The solution also confirms the general expectation
about the issue of time in quantum theory in the asymptotically flat
context. The parameter $t$ arises as the affine parameter along the
Hamiltonian vector field on the classical phase space and has the
space-time interpretation of time in the 3-metric defined by any
dynamical trajectory in the phase space. (This is also the situation
in full general relativity.) In the quantum theory, an analogous
parameter enters the Schr\"odinger equation (\ref{se}). However, since
general quantum states do not correspond to classical space-times,
this parameter does not have the standard interpretation of
time. This interpretation emerges only in the semi-classical
regime: in any coherent state, the parameter can be identified with
the classical $t$. Finally, we saw that the regulated metric and
holonomy operators can be constructed by a careful smearing procedure
which smoothens the sharp boundaries that enter the definition of their
classical analogs. The associated functional analysis subtleties are
non-trivial even from the mathematical perspective of a free field in
Minkowski space.

In the technical discussion, we made a liberal use of the fictitious
Minkowskian background $g^o_{ab}$ and the associated time parameter
$T$. However, this was done primarily for pedagogical reasons, i.e.,
to bring out the relation between the final quantum theory and the
expected one. We could have arrived at our Hilbert space of states
directly from the reduced phase space either by making use of the
``reality condition'' strategy \cite{aa:91} or by making an appeal to
null infinity and the S-matrix theory, without having to explicitly
introduce $g^o_{ab}$.

How do these results compare with those available in the literature?
Our analysis is closely related to that of Refs \cite{kk71, ma87}. In
the classical theory, the main difference lies in our systematic
handling of the asymptotically flat boundary conditions. In
particular, in our treatment, the true Hamiltonian arose directly from
the boundary term in the action. This point could not have been
realized in the early analyses because the relation between 3+1 and
2+1-dimensional theories was not well-known and, more importantly,
because a clear understanding of asymptotic flatness in 2+1 dimensions
has emerged only recently.  (Indeed, given what was known in the early
seventies, the treatment of Ref \cite{kk71} seems to be surprisingly
ahead of its time!)  In the quantum theory, the difference lies in the
treatment of certain functional analytical subtleties. That it is
necessary to regularize the metric operator was realized in
\cite{ma87}. However, the suggestion there that the softening of the
sharp boundaries can be brought about by a simple ultra-violet cut off
in the momentum space is incorrect; one needs suitable smearing fields
{\it in space-time}. Thus, our regularization differs from that in
\cite{ma87}.  Finally, our isolation of true degrees of freedom was
carried out in 2+1 dimensions. When translated to a 3+1 dimensional
perspective, our result is equivalent to the definition of true
observables given in \cite{ct91}.

Since the model has been solved exactly within the standard canonical
framework, it opens doors for further analysis in a number of
directions. We will conclude by mentioning a few of these.

First, we can now explore quantum field theory on a {\it quantum}
geometry. Part of the motivation here is similar to that of quantum
field theory in curved space-times; one wishes to investigate the
effects of a non-trivial background geometry on quantum
fields. Furthermore, this analysis can also shed light on the nature
of quantum geometry itself.  For instance, we may choose as our
background, a coherent state. The geometry influences the dynamical
evolution of the quantum field because the metric appears in the
expression of the Hamiltonian of the test field. Now, in the quantum
theory, we have two alternatives. First, we can consider just the
quadratic form that is determined by the (normal-ordered, unsmeared)
metric (\ref{mo}) and substitute its value in a coherent state in the
expression of the Hamiltonian. Since this value is just the classical
metric, this would lead us just to the standard quantum field theory
in curved space-times. To probe the effects of the {\it quantum}
nature of geometry, we would have to look beyond just the expectation
values. This can not be handled by a quadratic form alone; we need a
genuine operator. Thus, the second --and much more interesting--
possibility is to use the smeared metric operator in the expression of
the Hamiltonian of the test field. Then, one would see the effects of
the quantum geometry on the evolution of the matter, even in the case
when the geometry is assumed to be in the vacuum state
(initially). This analysis would be interesting because much of the
standard apparatus of quantum field theory in curved space-times uses
the fixed causal structure of the classcial geometry which is now
absent. Using the canonical framework, one would be able to do quantum
theory of test fields even when the causal structure is subject to
quantum fluctuations of its own.

Recall that, in the regularization of the metric operator, we needed a
smearing function $f_R$.  There is, however, no ``canonical'' choice;
while we know what the qualitative behavior of $f_R$ should be, there
is considerable freedom in its detailed form. Thus, we do not have a
``canonical'' regularized metric operator. All choices provide the
required ultra-violet cut-offs but the precise damping depends on the
specific form of $f_R$. The differences will show up, for example, in
the evolution of test fields.  It would be interesting to investigate
these differences and see if one can restrict the choice of the
smearing functions through thought experiments. If one can not, there
would a genuine quantization ambiguity. The situation would be similar
to that in non-relativistic quantum mechanics where, in general, the
factor ordering ambiguities can not be resolved purely on theoretical
grounds.

We saw that the regularized operators corresponding to the traces of
holonomies of connections are well-defined on the quantum Hilbert
space. Now, in the approach to quantum gravity based on these
holonomies \cite{aa:91}, a striking picture of quantum geometry has
emerged in which goemetrical operators such as areas and volumes have
a {\it discrete} spectrum. It is therefore natural to ask if the same
is true in the present case. The question is now manageable, thanks to
the regularized metric operators. Since the basic operator $:\!\hat
\gamma \! :$ is the regularized version of the restriction of the
Hamiltonian in a box, it is quite likely that its spectrum is
discrete. If so, the lengths in the radial directions and areas will
be quantized. This would be a striking result coming from a Fock-like
representation.

Another direction for further investigation is provided by the Gowdy
models. Since these are spatially compact and have initial curvature
singularities, new issues arise. These will be discussed in the sequel
to this paper. While both these problems deal only with the ``one
polarization'' case  --the two Killing fields are hyper-surface
orthogonal in the 3+1-dimensional picture--  one can also investigate
the two polarization case \cite{bbetal94}. In the case when the
translational Killing field is time-like, this case was analyzed in
detail by Korotkin and Nicolai \cite{dkhn96} recently. Their
quantization is mathematically complete but somewhat unconventional in
the sense that the relation between their Hamiltonian description and
the standard Poisson-brackets of classical general relativity is
unclear. It would be interesting to compare the results obtained here
with the reduction of their model to the one polarization case. More
recently, infinite number of conserved quantities have been
constructed in the classical theory with two polarizations
\cite{aavh96}. Using these, one may be able to extract the true
degrees of freedom in this more general case and quantize the model
along the lines of this paper.

Finally, the present model itself offers an attractive setting to
explore the idea of ``fuzzing'' of space-time points using techniques
involving null infinity \cite{tnetal96}. As mentioned in section
\ref{s4.1}, the 2+1-dimensional space-times considered here are
asymptotically flat at null infinity \cite{aajbbs}. Furthermore, since
the form of the metric is sufficiently simple, it should be possible
to integrate the null geodesics and express the ``light cone cuts of
null infinity'' explicitly in terms of the initial data for the scalar
field at null infinity. These cuts label space-time points. The
asymptotic quantization of the scalar field \cite{aa:aq}  --which is
equivalent to the quantization presented here--  would then lead to
fuzzy points.  So far, in this approach, detailed calculations have
been carried out only in the linearized approximation \cite{tnetal96}.
The underlying simplicity of cylindrical waves provides an interesting
arena where these results can be extended beyond the linear context.

\vskip.5cm
\ni
{\Large\bf  Acknowledgments}
\vskip.3cm 
We are grateful to Curt Cutler, Ted Newman, Jorge Pullin and
especially Madhavan Varadarajan for valuable discussions.
%and especially for suggesting the second gauge condition in
%(\ref{gfc}).
We would also like to thank Arthur Jaffe and Thomas Thiemann for
pointing out to us Ref \cite{cj70} and \cite{rs} respectively.  This
work was supported in part by the NSF grant 95-14240, by the Eberly
Research Funds of the Pennsylvania State University, and by a
Fellowship from Capes (Brazil) to MP.

\end{document}